\documentclass[aps,pra,reprint,groupedaddress]{revtex4-1}
\usepackage{amsmath,amssymb}
\usepackage{graphicx}
\usepackage{units}
\usepackage{latexsym}
\usepackage{bm}
\usepackage{color}
\usepackage{epstopdf}
\usepackage{enumitem}
\graphicspath{ {./figures/} }

\newcommand{\abs}[1]{\left\lvert#1\right\rvert} 
\begin{document}
\title{\large Multiphoton Double Ionization of Helium at 394\,nm - a Fully Differential Experiment }
\author{
K. Henrichs$^1$, S. Eckart$^1$, A. Hartung$^1$, D. Trabert$^1$, K. Fehre$^1$, J. Rist$^1$, H. Sann$^1$, M. Pitzer$^{2,3}$,\\ M. Richter$^1$, H. Kang$^{1,4}$, M. S. Sch\"offler$^1$, M. Kunitski$^1$, T. Jahnke$^1$} \author{R. D\"orner$^1$}\email{doerner@atom.uni-frankfurt.de} 
\affiliation{$^1$Institut f\"ur Kernphysik, J.~W.~Goethe-Universit\"at, Max-von-Laue-Str. 1, 60438 Frankfurt am Main, Germany\\
$^2$Institut f\"ur Physik, Universit\"at Kassel, Heinr.-Plett-Str. 40, 34132 Kassel, Germany\\
$^3$Institute of Chemical and Biological Physics, Weizmann Institute of Science, P.O. Box 26, 76100 Rehovot, Isreal\\
$^4$State Key Laboratory of Magnetic Resonance and Atomic and Molecular Physics, Wuhan Institute of Physics and Mathematics, Chinese Academy of Sciences, Wuhan 430071, China}
\date{\today}
\begin{abstract}
{We report on a kinematically complete experiment on strong field double ionization of helium using laser pulses with a wavelength of \unit{394}{\,nm} and intensities of $\unit{3.5-5.7\times10^{14}}{\,W/cm^2}$. Our experiment reaches the most complete level of detail which previously has only been reached for single photon double ionization. We give an overview over the observables on many levels of integration, starting from the ratio of double to single ionization, the individual electron and ion momentum distributions over joint momentum and energy distributions to fully differential cross sections showing the correlated angular momentum distributions. Within the studied intensity range the ratio of double to single ionization changes from $2\times 10^{-4}$ to $1.5\times 10^{-3}$. We find the momentum distributions of the $\rm{He}^{2+}$ ions and the correlated two electron momentum distributions to vary substantially. Only at the highest intensity both electrons are emitted to the same direction while at the lowest intensity back-to-back emission dominates. The joint energy distribution of the electrons shows discrete structures from the energy quantization of the photon field which allows us to count the number of absorbed photons and thus access the parity of the final state. We find the energy of the individual electron to show a peak structure indicating a quantized sharing of the overall energy absorbed from the field. The joint angular momentum distributions of the two electrons show a highly directed emission of both electrons along the polarization axis as well as clear imprints of electron repulsion. They strongly change with the energy sharing between the electrons. The aspect of selection rules in double ionization which are also visible in the presented dataset has been subject to a preceding publication \cite{Henrichs18prar}.}
\end{abstract}
\maketitle
\section{Introduction}
Double ionization of Helium has been the showcase process to study electron correlation using electron impact \cite{Dorn01prl}, ion impact \cite{Moshammer96prl,Schulz2005jpb}, antiparticle impact \cite{Andersen86}, single photon absorption \cite{Schwarzkopf93prl,Briggs00jpba,Doerner96fullydiff}, Compton scattering \cite{Spielberger95prl,Spielberger96prl}, FEL radiation \cite{Rudenko95prl} and femtosecond laser pulses (see \cite{Becker12rmpa,Becker05jpbreview,Doerner02advancesa} for reviews). The mechanisms leading to ejection of the two electrons can roughly be grouped in those relying on electron-electron correlation and those which would exist even in the absence of electron-electron interaction. Different communities have given different names to these mechanisms. Correlation driven double ionization is called non-sequential in the laser community, while studies with other projectiles have identified correlation mediated mechanisms as shake-off, TS1/knock-off \cite{Knapp02prl} and the Quasi-free mechanism \cite{Schoeffler13prl}. The correlation free double ionization is commonly referred to as sequential double ionization \cite{Becker12rmpa} in the laser community and TS2 \cite{McGuirebook} in charged particle impact studies. The latter processes are absent for single photon absorption and Compton scattering. Helium is the paradigmatic target in all these fields of atomic physics mainly because of its simplicity and theoretical tractability.

The fundamental nature of double ionization and the fact that only three particles (four for particle impact) are involved make it a worthy and in principle achievable goal to reach the ultimate level of detail in experimental studies. This implies to measure fully differential cross sections, e.g. to avoid integration over any unobserved quantity. However, this ultimate goal has only been reached in the most mature fields of electron impact \cite{Dorn01prl} and single photon impact \cite{Briggs00jpba} studies. It is the goal of the present paper to show the first fully differential experiment for multi-photon strong field double ionization. One special aspect, the quantum mechanical selection rules, has been presented in a preceding report \cite{Henrichs18prar}. In the multi-photon context fully differential rates entail that the number of photons and the momentum vectors of two of the three particles (ion and two electrons) in the final states are measured. The momentum of the third particle then is fixed by momentum conservation. Within the dipole approximation for linearly polarized light the cylinder symmetry around the polarization vector reduces the dimensionality of the problem to five dimensions.    

In our experiment we use laser pulses of $394$\,\unit{nm} wavelength ($h\nu=3.15$\,\unit{eV}) at 40\,\unit{fs} with intensities of $3.5$, $4.6$ and $5.7$\,$\times 10^{14}W/cm^2$. This places our study between the much discussed cases of two photon double ionization in the perturbative regime \cite{Rudenko95prl} and long wavelength tunneling regime \cite{Becker12rmpa}. At the $394$-nm wavelength the ponderomotive energy ($U_p$) of the electron in the field is $5.1$ ($6.7$, $8.3$)\,\unit{eV} at the intensities of  $3.5 \times 10^{14}W/cm^2$ ($4.6$, $5.7$\,$\times 10^{14}W/cm^2)$ studied in this work. Thus the maximum energy of 3.17 $U_p$ with which an electron can reencounter its parent ion is $26$\,\unit{eV}, much below the second ionization potential of Helium ($54$\,\unit{eV}). This together with the short wavelength limits the validity of the classical rescattering model in our case. Nevertheless, we will refer to the simple rescattering scenario to place our results in the strong field context. 

We have used $394$\,\unit{nm} instead of a longer wavelength, because we aimed to determine the number of photons absorbed in each ionization event. The discrete nature of the photon field leads to discrete structures in the electron energy distribution for single ionization at:
\begin{eqnarray}
	E_1 = n h\nu - I_{p1} - U_p~.
\label{eqneesingle}
\end{eqnarray}
where $I_{p1}$ refers to the first ionization potential of the atom ($24.6$\,\unit{eV} for He). For double ionization similar peaks arise in the electron's sum energy ($E_1+E_2$) \cite{Lein01pra} which we refer to as ATDI (above-threshold-double-ionization) peaks at
\begin{eqnarray}
	E_1 + E_2 = n h\nu - I_{p1} - I_{p2} - 2U_p~.
\label{eqneedouble}
\end{eqnarray}
where $I_{p2}$ refers to the second ionization potential of the atom ($54$\,\unit{eV} for He). These peaks 
become experimentally unobservable due to volume averaging at larger $U_p$.

The ATDI peaks have been predicted already based on solutions of the time-dependent Schr\"odinger equation (TDSE) in two dimensions and later been confirmed in many calculations \cite{Lein01pra,Liao10pra,Parker06prl,Zielinski16pra,Thumm14pra}. 

The paper is organized as follows. We briefly describe the relevant details of our experiment using the COLTRIMS (cold-target recoil-ion momentum spectroscopy) technique and our intensity calibration in section \ref{sectionexp}. We then show some results for single ionization (section \ref{sectionsingle}). For double ionization we start with global observables such as ratios of double to single ionization (section \ref{chapterratio}), move on to electron and ion momentum distribution as well as energy distributions (section \ref{chapterenergy}) and then present joint electron energy distributions (section \ref{sectionjoint}). Finally we discuss fully differential rates (section \ref{chapterfully}). We close with some conclusions.

\section{Experiment \label{sectionexp}}
\subsection{Laser and COLTRIMS Reaction Microscope } 
In our experiment, we used linearly polarized femtosecond laser pulses at a wavelength of $394$\,\unit{nm} ($h\nu\approx 3.15$\,\unit{eV})  and employed a COLTRIMS Reaction Microscope \cite{Ullrich03rpp,Jahnke04jesp,Doerner00pr} for the coincidence detection of all charged particles. A Ti:Sa laser system (Wyvern-500, KMLabs, $45$\,\unit{fs}, $100$\,\unit{kHz} at a central wavelength of $788$\,\unit{nm} was used to generate the second harmonic at $394$\,\unit{nm} with a $200$\,$\mu m$  BBO crystal. The laser pulse was backfocussed by a spherical mirror of $60$\,\unit{mm} focal length into a supersonic gas-jet. 
As we use the momentum vector of the ion in part of our analysis it was essential to achieve a narrow momentum distributions of the atoms in all three dimensions. To this end helium gas was precooled to 30 Kelvin and expanded through a $5$\,$\mu m$ nozzle at a driving pressure of $6$\,bar into vacuum. This resulted in a speed ratio of about 200 \cite{Brusdeylins89}. About $5$\,\unit{mm} downstream of the nozzle the jet passed a $0.3$\,\unit{mm} diameter skimmer entering a differential pumping stage of $27$\,\unit{mm} length. This pumping stage was separated by a $0.15$\,\unit{mm} diameter aperture from the main experimental chamber housing the spectrometer. The distance from the jet inlet aperture to the laser focus in the center of the spectrometer was about $60$\,\unit{mm}. In the differential pumping stage right before the aperture the gas jet could be collimated by razor blades which were mounted on piezo actuators and could be moved with nm precision. With these collimators the gas-jet was cut to intersect only with the central part of the laser focus to reduce focal averaging and to adjust the count rates.

The ion arm of the spectrometer consisted of an acceleration region ($18.2$\,\unit{cm}) with an electric field of ${2.08}\,\unit{V/cm}$ followed by a $40$\,\unit{cm} field free drift region. The electron arm consisted of a $7.8$\,\unit{cm} acceleration region and no drift region. A homogeneous magnetic field ($7.2$\,G) parallel to the electric field was used to guide electrons towards the detector. Typical count rates during the experiment were $7$\,\unit{kHz} ions (including residual gas) and $20$\,\unit{kHz} electrons at $100$\,\unit{kHz} laser repetition rate. Ions and electrons were detected by $80$\,\unit{mm} active area micro channel plate detectors equipped with a hexagonal delay line anode for multiple hit position readout \cite{Jagutzki02ieee}. 

\begin {figure}[t]
  \begin{center}
    \includegraphics[width=0.9\linewidth]{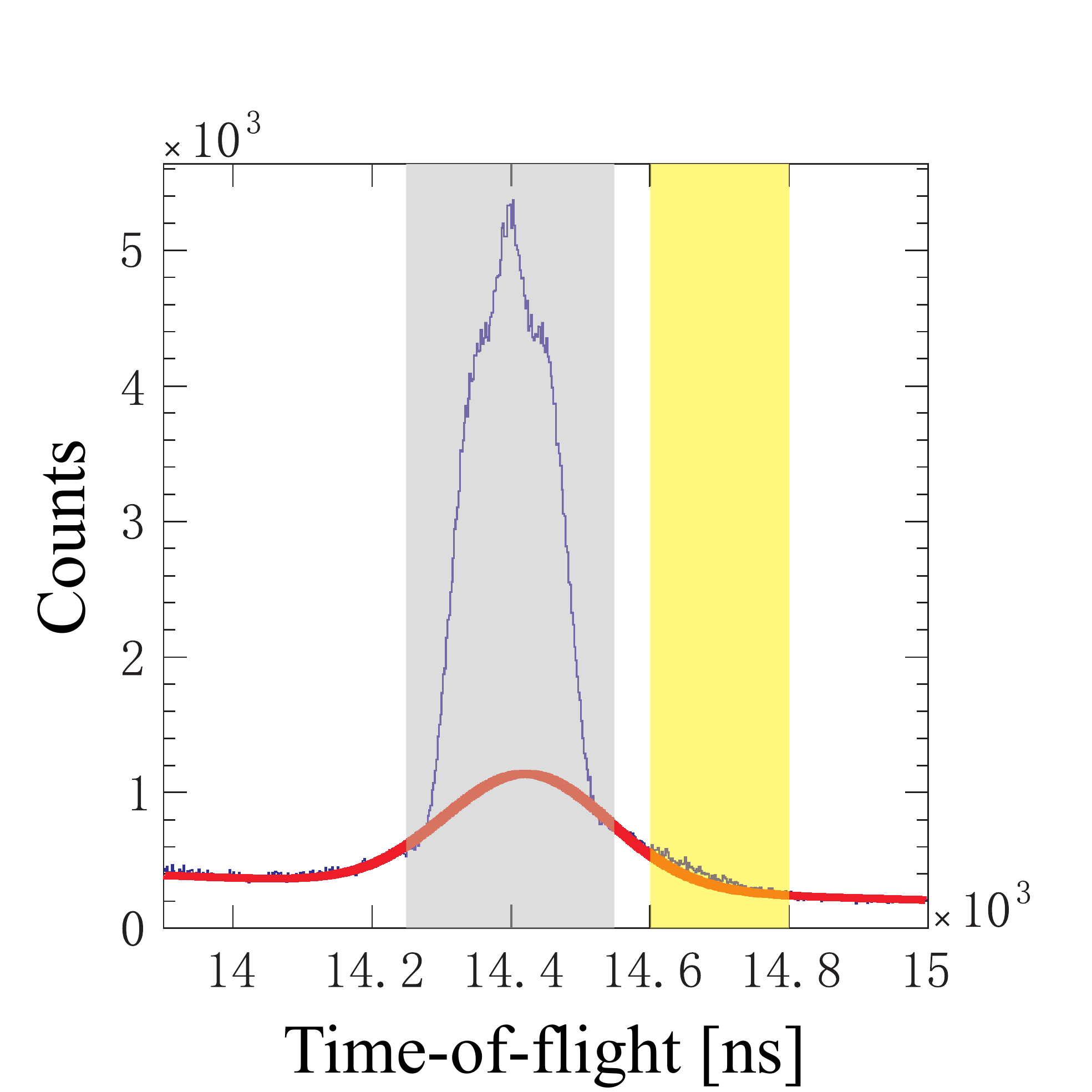}
\caption{Ion time-of-flight distribution  gated on momentum components $< 0.8$\,a.u. in both directions perpendicular to the spectrometer axis ($\rm{He}^{2+}$ and $H_2^{+}$ are expected at a TOF of $14.4$\,\unit{$\mu$s}). The red curve indicates background from thermal $H_2^+$ ions from the residual gas. Ions in the yellow region are used for background subtraction.}   
  \label{figh2back}
  \end{center}
\end {figure}

In the interaction chamber the background pressure was below $2 \times 10^{-11}$~mbar. This was essential to reduce the amount of $H_2^+$ ions from ionization of residual gas which overlaps in time-of-flight (TOF) with the $\rm{He}^{2+}$ ions. Since the residual gas is at room temperature, the $H_2^+$ ions have a broad momentum distribution. In Fig. \ref{figh2back} we show the measured TOF distribution in the mass region of $H_2^+ / He^{2+}$ in which we have already selected events with small momentum components in the two dimensions perpendicular to the field direction of the spectrometer (which coincides with the polarization direction of the laser field). For the part of the data analysis where we have used events in which one electron has been detected and the momentum of the second electron is inferred using the ion momentum and momentum conservation, we have subtracted a background from $H_2^+$ as indicated in Fig. \ref{figh2back}.

\subsection{Intensity calibration} 
\label{sectionintensity}
The primary calibration of our intensity was done by measuring the energy shift of the ATI peaks of Xe as function of pulse energy (Eq. \ref{eqneesingle}). This establishes a proportionality constant between the $U_p$ determined in situ (in the focus) and the energy-per-pulse of our laser measured outside the vacuum vessel. The intensity calibration obtained this way has been cross-checked in the Helium data by inspecting the cutoff in the single ionization spectra and the energy of the higher Helium ATI peaks (see Fig. \ref{figeeinfach}). A further in situ cross-check on the actual data itself is the location of the maxima of the ATDI peaks which shifts sensitively with $U_p$ according to equation \ref{eqneedouble} (see Fig. \ref{figeesum}).  We note that in \cite{Henrichs18prar} we have shown data from the lowest intensity reported here. In \cite{Henrichs18prar} the intensity was indicated to be $3\times10^{14}$. After careful recalibration of the intensity we now believe that the intensity for this measurement is $3.5\times10^{14}$. This does not alter any conclusions in \cite{Henrichs18prar}.  We estimate that our recallibrated intensities given in table \ref{table} are accurate to better than 20\%.

\begin{table}[htbp]
	\centering
		\begin{tabular}{ccccc}
			\hline
    \textbf{Intensity} & \textbf{ $U_p$}  & \textbf{$E_{return}^{max}$} & \textbf{ $2\sqrt{U_p}$} & \textbf{Keldysh }\\			
    \textbf{$\left[W/cm^2\right]$} & \textbf{ $\left[eV\right]$} & \textbf{ $\left[eV\right]$} & $\left[a.u.\right]$ & \textbf{Parameter} \\ \hline
    $3.5\times10^{14}$ & 5.1 & 16 & 0.86 & 1.6 \\\hline
    $4.6\times10^{14}$ & 6.7 & 21 & 0.99 & 1.4 \\\hline
    $5.7\times10^{14}$ & 8.3 & 26 & 1.10 & 1.2 \\\hline
 		\end{tabular}
		\caption{Characteristics of laser intensities used in the current work ($394$\,\unit{nm})}
		  \label{table}
\end{table}

\section{Single ionization  \label{sectionsingle}}
\begin {figure}[t]
  \begin{center}
    \includegraphics[width=0.9\linewidth]{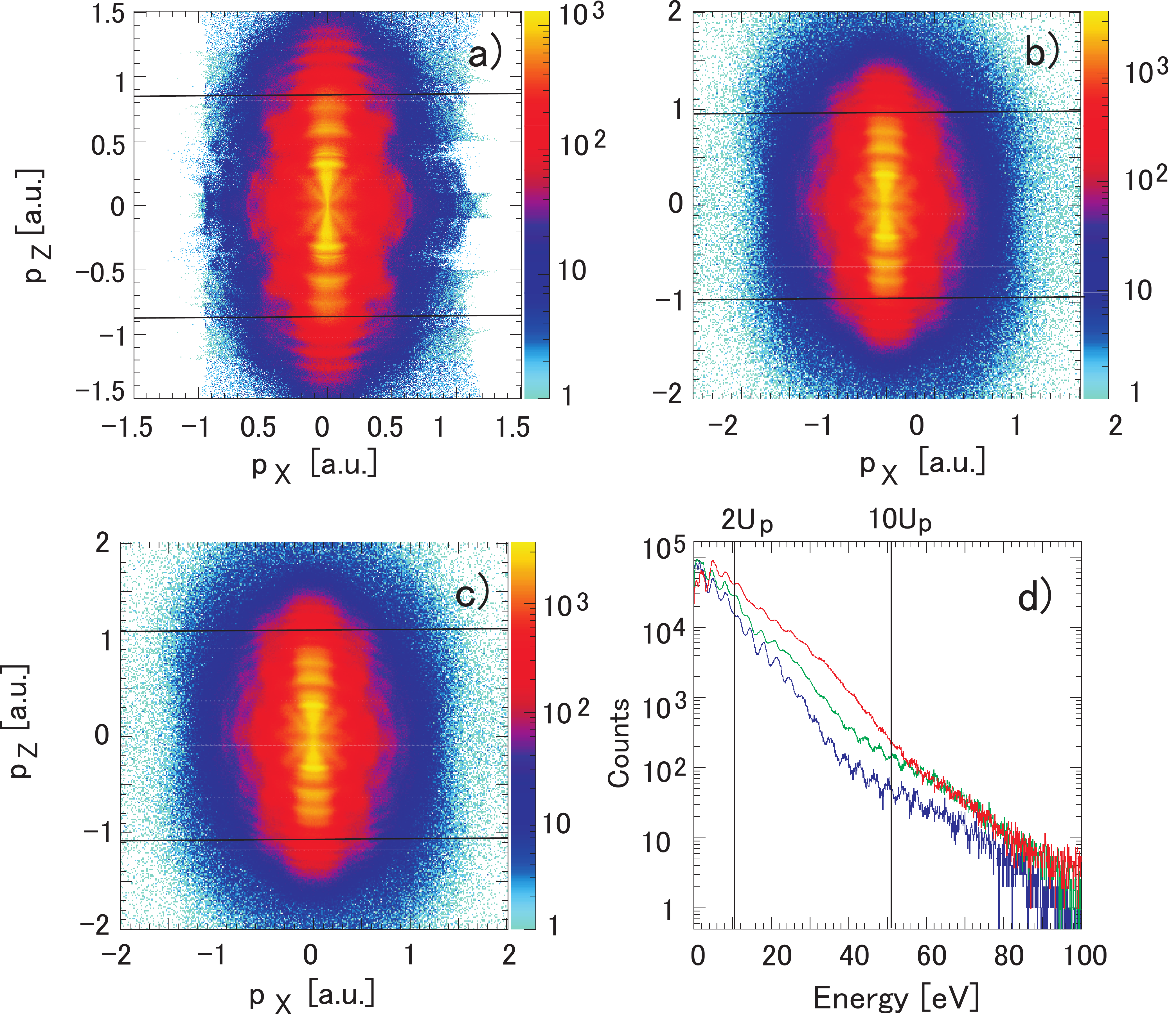}
    \caption{Electron momentum and energy distributions from single ionization of He at $394$\,\unit{nm} (a) $ 3.5\times 10^{14}$\,$W/cm^2$, (b) $ 4.6\times 10^{14}$\,$W/cm^2$, (c) $5.7\times 10^{14}$\,$W/cm^2$. In (a)-(c) the black lines indicates the maximum classical momentum of $2\sqrt{U_p}$. The laser's polarization is aligned along the z-axis. (d) same data as (a-c) but converted to electron energy. In d) the lines showing $2U_p$ and $10U_p$ refer to the lowest intensity.}
  \label{figeeinfach}
  \end{center}
\end {figure}

\begin {figure}[t]
  \begin{center}
    \includegraphics[width=1.0\linewidth]{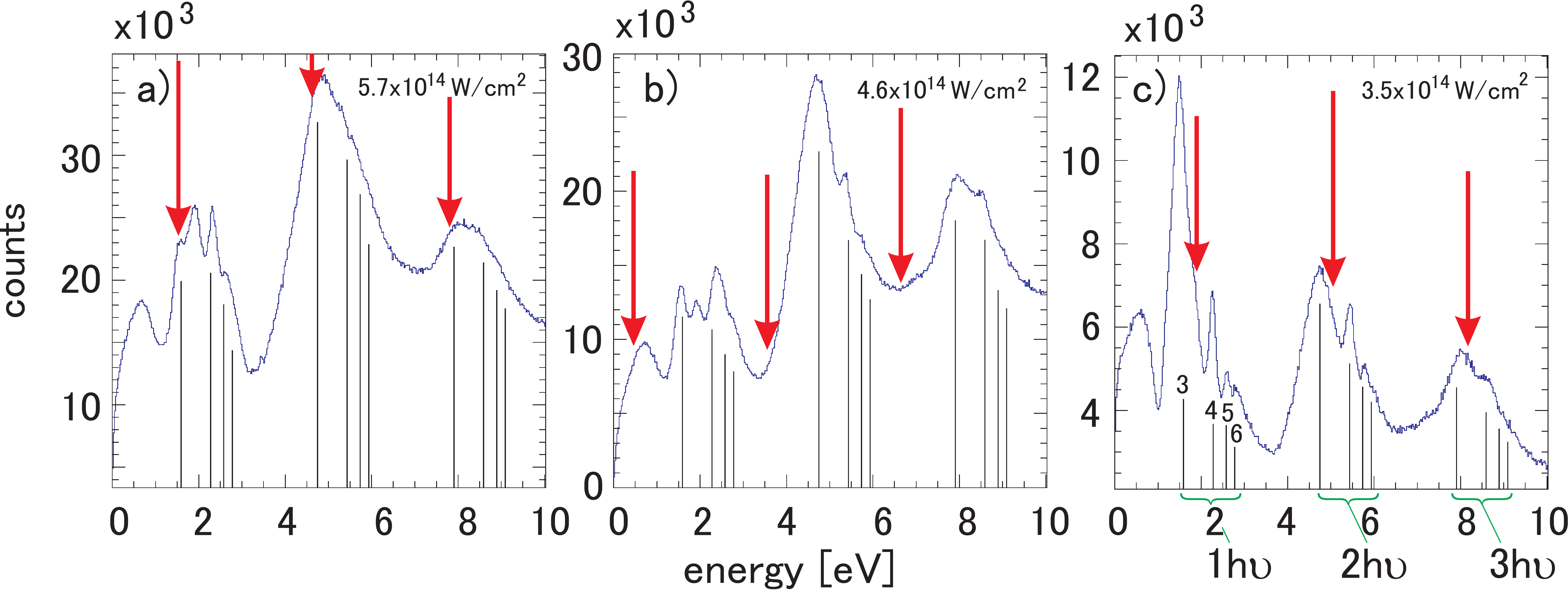}
    \caption{Same data as Fig. \ref{figeeinfach}, zoomed into the low-energy region. The red arrows mark the expected locations of ATI peaks (Eq. 1) while the black lines show the intensity independent energies at which Freeman-resonances are expected \cite{Freeman91jpb} (1-, 2- and 3- photon ionization of a resonantly populated Rydberg-state of the quantum number that is indicated in (c)).}
  \label{figefreeman}
  \end{center}
\end {figure}

For single ionization the electron momentum distribution (Fig. \ref{figeeinfach}) shows the features well known in this intensity regime. The distribution is concentrated along the polarization axis and shows side lobes at small momenta \cite{Richter15prl}. We have indicated the maximum momentum a free electron can acquire from acceleration in the field without rescattering by ($2\sqrt{U_p}$) (see table \ref{table}). A drop of intensity at this cutoff can be seen even on  logarithmic color scale. The energy distributions for the three intensities we studied (Fig. \ref{figeeinfach}b) show the well known increase of electron energies with intensity together with a plateau like feature between $2$\,$U_p$  and $10$\,$U_p$ at the highest intensity. At electron energies above $10$\,\unit{eV} a clear sequence of ATI peaks can be seen. At lower energies we observe multiple peaks which are not located at $E_e=nh\nu - I_{p1} - U_p $. These peaks do not shift in energy as the intensity is varied. Those peaks are known as Freeman resonances \cite{Freeman91jpb} and result from single (two, three) photon ionization of Rydberg-states which are populated resonantly on the rising edge of the pulse. The progression of these Rydberg-states is shown by the lines in Fig. \ref{figefreeman}.

\section{Double ionization}
\subsection{Ratio of double to single ionization \label{chapterratio}}
The most global observable allowing to characterize the double ionization mechanism is the ratio $R$ of double to single ionization as function of intensity. The pioneering work of \cite{Walker94prl} for $800$\,\unit{nm} found a knee-shaped feature. Over the intensity range of this knee, R is almost intensity independent, indicating that the second electron is removed by excitation or knock-off upon rescattering of the primary electron. At about the intensity at which the maximum classical return energy of the primary electron falls below the threshold for electron impact excitation ($40.8$\,\unit{eV}) or ionization ($54$\,\unit{eV}) of the $\rm{He}^+$, $R$ drops steeply for $800$\,\unit{nm} (see \cite{Eckart16prl,Xie14prl,Mancuso16prl,Bergues12natcom} for other means to control the return energy). 

\begin {figure}[t]
  \begin{center}
    \includegraphics[width=0.9\linewidth]{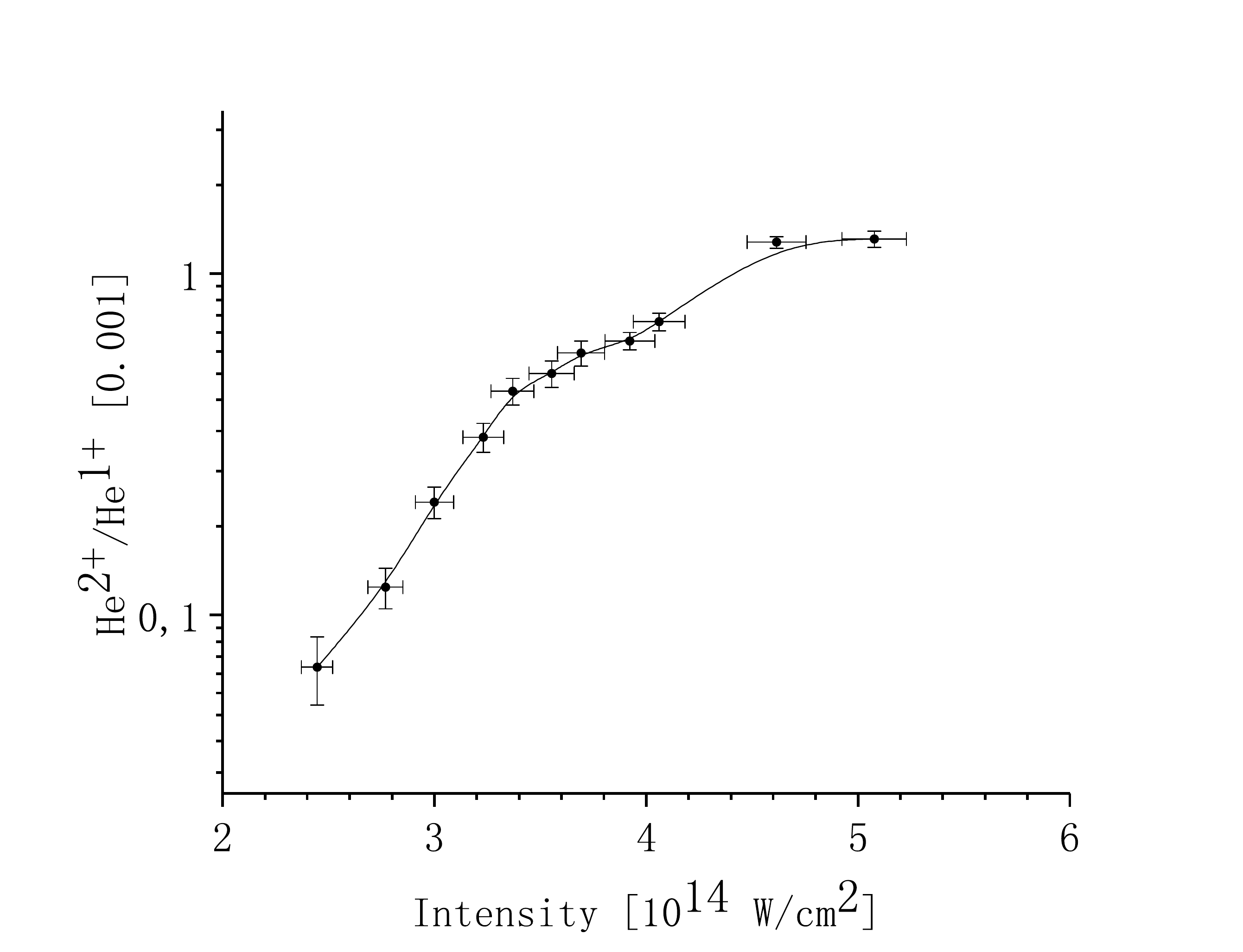}
    \caption{Ratio of $\rm{He}^{2+}/He^{1+}$ as function of intensity for a wavelength of $394$\,\unit{nm} at a pulse length of $45$\,\unit{fs}. The gas target was narrower than the Rayleigh length of the focus (see text).}
  \label{figratio}
  \end{center}
\end {figure}

For $394$\,\unit{nm} a similar knee structure has been reported in theory and experiment (see \cite{Chen18pra} for a recent collection of experimental and theoretical data). Fig. \ref{figratio} shows that our data are taken at an intensity at the steep drop of $R$ (below the knee). We note that the absolute value of our measured $R$ is consistent with \cite{Sheehy98pra}, however, our data are shifted to much lower intensities. Our pulse length is $45$\,\unit{fs} as compared to $120$\,\unit{fs} in \cite{Sheehy98pra}. Further the data in \cite{Sheehy98pra} are volume averaged while we have at least partially avoided the volume averaging by limiting our gas target to be much narrower than the Rayleigh length. All of this does not fully explain the discrepancy between \cite{Sheehy98pra} and our intensities. We emphasize that we have measured the intensity in situ and cross checked this finding in the highly differential data (as explained in section \ref{sectionintensity}). Furthermore we have also measured $R$ for a Ne target and found excellent agreement with the data of \cite{Sheehy98pra} as reported in \cite{Chen18pra} (not shown).   

\subsection{Electron and ion momentum distributions \label{chapterenergy}}
For double ionization the sum momentum of both electrons is given (neglecting the photon momentum) by the momentum of the doubly charged ion. These ion momentum distributions are highly informative. For double ionization by $800$\,\unit{nm} light they allowed to clearly prove rescattering to be responsible for double ionization \cite{Weber00prl,Moshammer00prl,Weber00jpb,Becker00prl,Kang18prl}. The maximum momentum that a doubly charged particle can acquire from acceleration by a non-relativistic laser field is given by $4\sqrt{U_p}$. 

This momentum corresponds to charging the particle at the zero crossing of the electric field. In this case it receives a momentum given by the charge times the negative vector potential. In the rescattering scenario a first electron escapes and is driven by the field. It will have the maximum possible recollision energy of $3.17$\,$U_p$ when it recollides with its parent ion at the zero crossing of the field. If the recollision energy is equal to the second ionization potential this leads to both electrons and the ion being almost at rest at the time of the field's zero crossing. Such a scenario thus results in a double hump structure of the ion momentum distribution at $\abs{p^{2+}_{z}}=4\sqrt{U_p}$ where $p^{2+}_{z}$ is the momentum of the doubly charged ion along the polarization axis. If the recollision leads to excitation and the excited electron is freed later in the pulse, this leads to a filling up of the double hump structure at the origin \cite{Rudenko07prlb}. 

\begin {figure}[t]
  \begin{center}
    \includegraphics[width=1.0\linewidth]{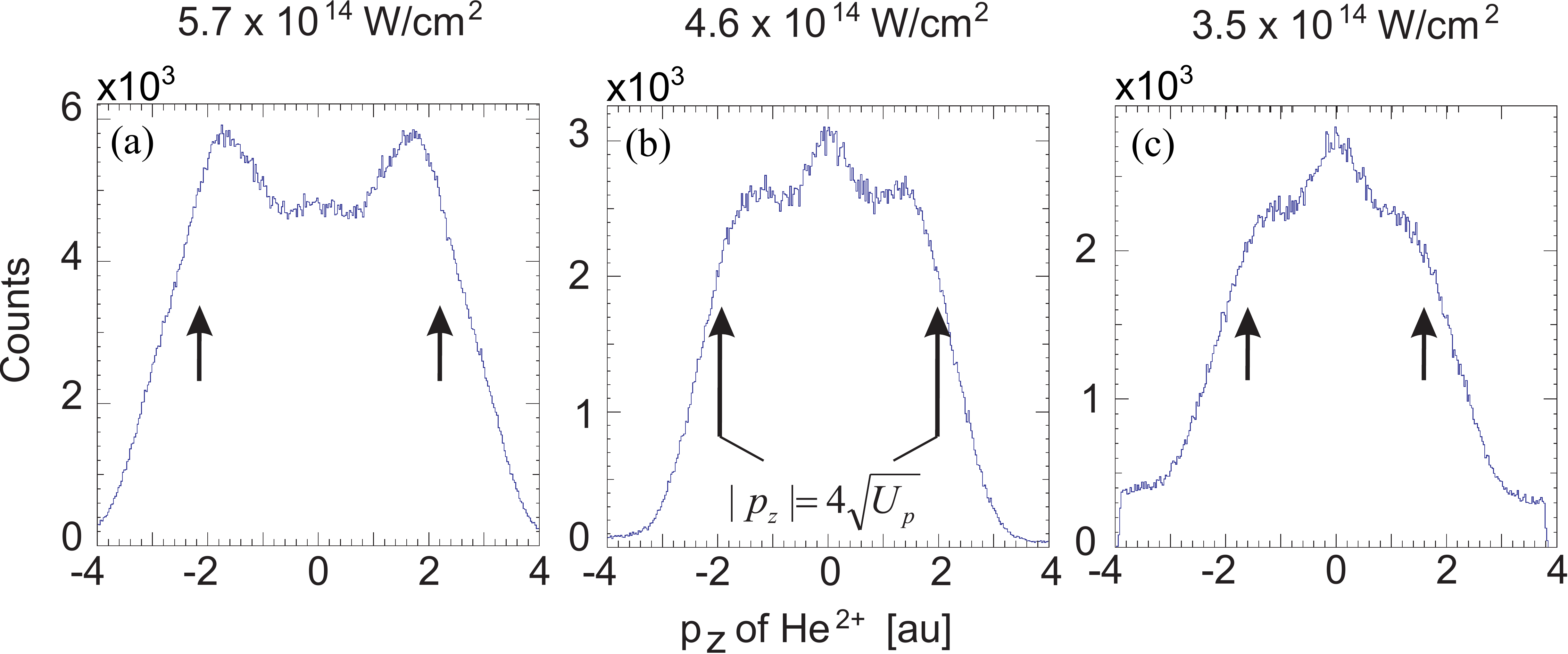}
    \caption{$\rm{He}^{2+}$ ion momentum distribution parallel to the light polarization for double ionization of He at $394$\,\unit{nm}. Intensities as given in the panels. The arrows indicate the momentum corresponding to twice the maximum vector potential of $4\sqrt{U_p}$, which is the momentum a double charged ion would receive, if it was born into the laser field a the zero crossing of the field. }
  \label{figpz2}
  \end{center}
\end {figure}

\begin {figure}[t]
  \begin{center}
    \includegraphics[width=1.0 \linewidth]{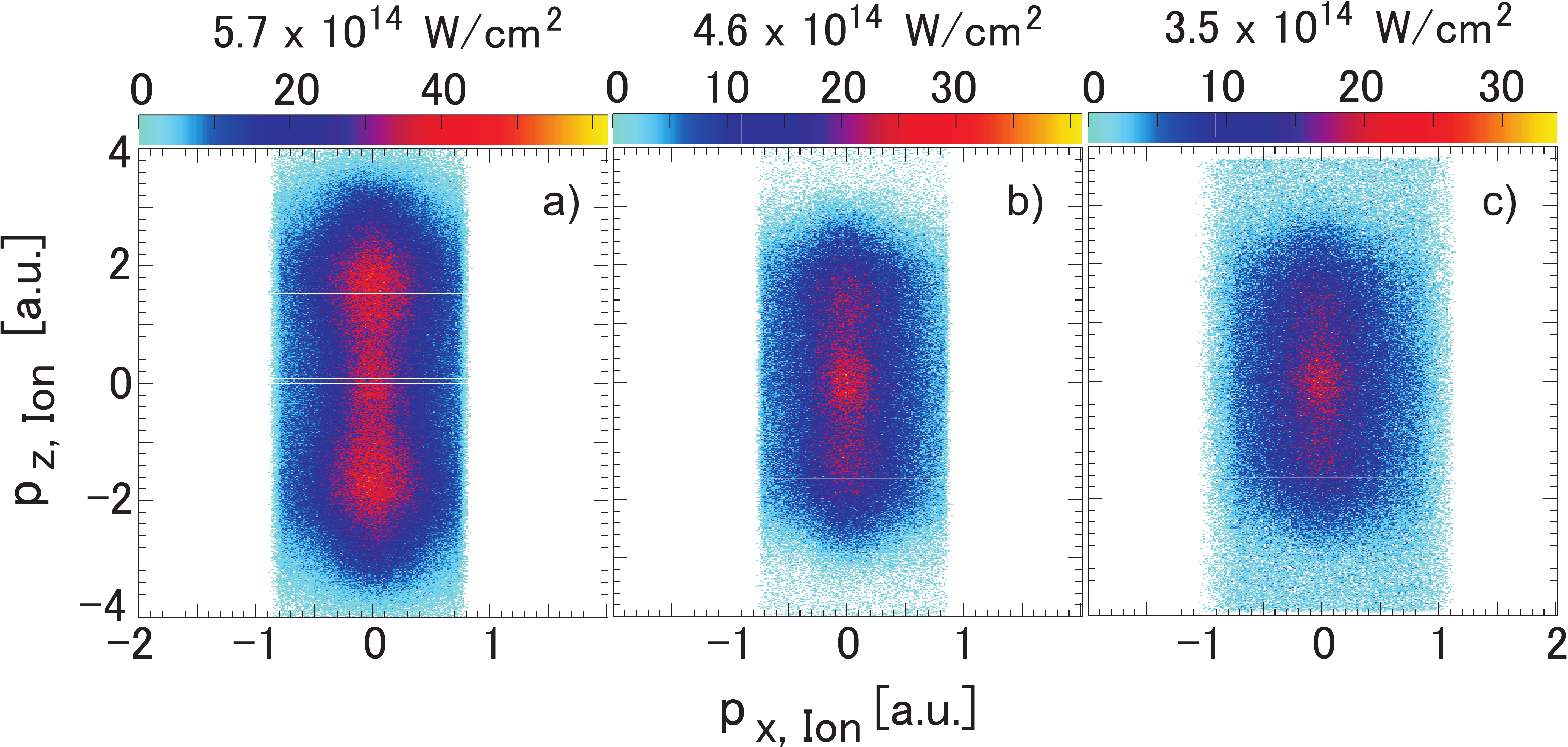}
    \caption{$\rm{He}^{2+}$ ion momentum distributions for double ionization of He at $394$\,\unit{nm}. Horizontal axis: one of the momentum components perpendicular to the light polarization,  vertical axis: momentum component parallel to the polarization. The third momentum component is integrated over. Fig.  \ref{figpz2} shows a projection of the data onto the vertical axis.}
  \label{fig2dimhe2}
  \end{center}
\end {figure}

In Fig. \ref{figpz2} we show the measured $\rm{He}^{2+}$ momentum distributions for double ionization of He at $394$\,\unit{nm}. At all three intensities the distributions extend to almost $4\sqrt{U_p}$. At the highest intensity we observe a clear double hump structure while at lower intensities a narrow peak at zero momentum fills the valley of the momentum distribution. Inspecting the two-dimensional ion momentum distributions (Fig. \ref{fig2dimhe2}) supports that there are two distinguishable features in the momentum distribution; a broad peak at large positive and negative momenta and a rather narrow feature at the origin.

\begin {figure}[t]
  \begin{center}
    \includegraphics[width=1.0 \linewidth]{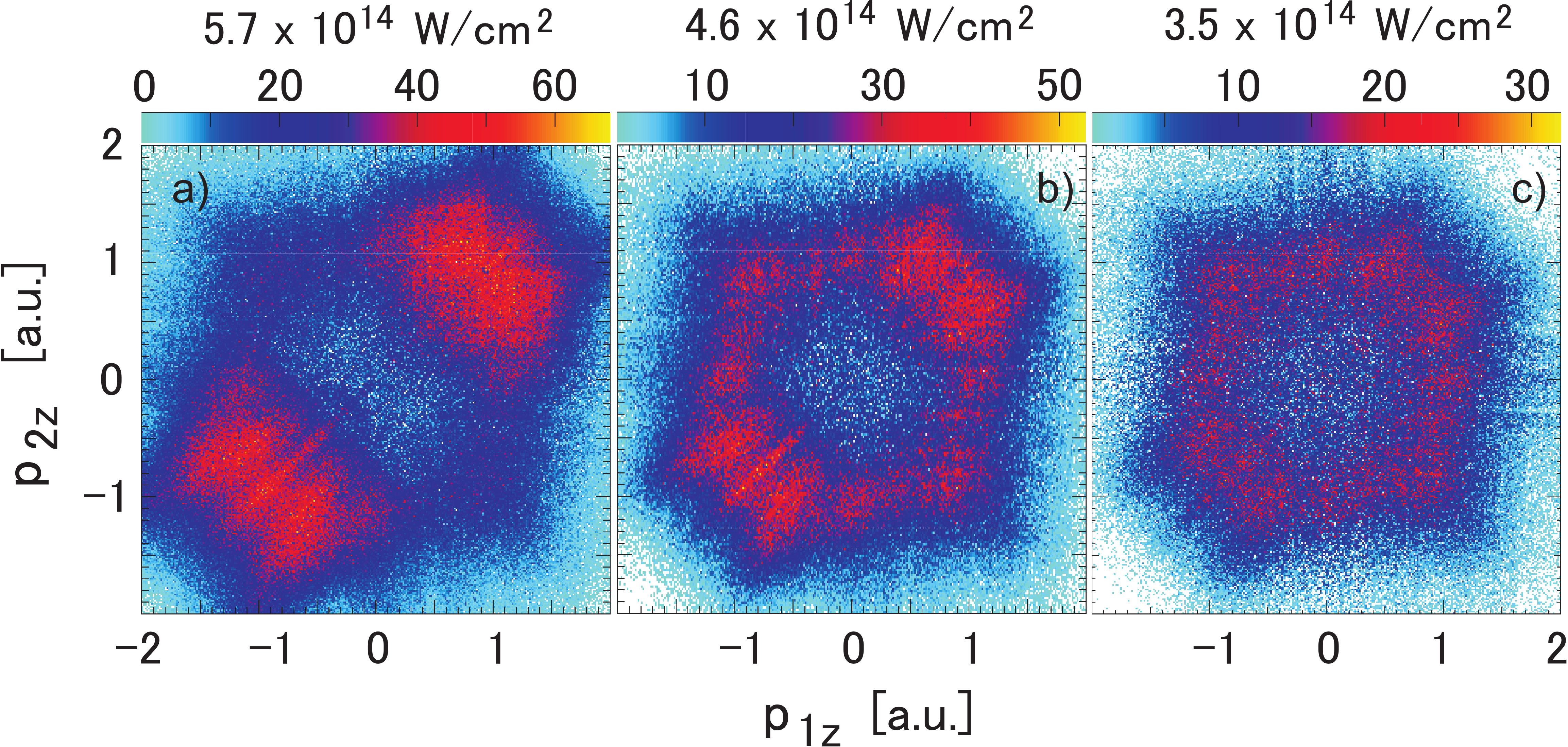}
    \caption{Joint momentum distributions of the two electrons for double ionization of He at $394$\,\unit{nm}. The horizontal (vertical) axis shows the momentum component of electron 1 (2) parallel to the light polarization axis. The data are integrated over all other momentum components. To avoid artifacts originating from the dead time of the detector, one electron and the ion is detected, the second electrons momentum is obtained using momentum conservation. Intensities as given in the panels.}
  \label{figpz1pz2}
  \end{center}
\end {figure}

To further illuminate the origin of this structure we plot in Fig. \ref{figpz1pz2} the momenta of both electrons parallel to the electric field. These joint parallel momentum distributions have been successfully used to unveil the double ionization mechanism \cite{Weber00nature,Weckenbrock03prl,Weckenbrock04prl,Staudte07prl,Rudenko07prlb,Bergues12natcom,Kang18pra}. It clearly shows a prominent feature in the first and third quadrant which indicates that both electrons are emitted side-by-side into the same hemisphere (both with similar momenta close to $2\sqrt{U_p}$). A second feature shows back-to-back emission. With increasing intensity the side-by-side emission dominates over back-to-back emission. This is very similar to what has been seen already for $800$\,\unit{nm} \cite{Liu08prl}. At all intensities the region close to the origin is almost empty of counts. The electron momenta along the polarization direction can to some extend be interpreted as ionization times, with time being mapped to momentum by the negative vector potential at the instant when the electron is set free. Therefore the fact that at least one of the two electrons has substantial momentum of $1\sqrt{U_p}$ to $2\sqrt{U_p}$ shows that this electron is set free around the zero crossing of the electric field (maximum of the vector potential). This in turn is a strong evidence that double ionization is induced by recollision.

Taking the low intensity into account this finding is not self-evident as the maximum recollision energy is only $16$\,\unit{eV} at our lowest and $26$\,\unit{eV} at the highest intensity. This is much below the threshold for electron impact ionization of $54$\,\unit{eV} and also below the first excitation threshold of $\rm{He}^+$ at $40.5$\,\unit{eV}. Thus the knock-off or excitation by recollision (RESI) requires the absorption of many additional photons upon recollision. 

The momentum distribution in the first and third quadrant of Fig. \ref{figpz1pz2} shows a minimum along the diagonal. This is in line with a similar observation at $800$\,\unit{nm} \cite{Staudte07prl,Rudenko07prlb} and was taken as evidence for electron repulsion \cite{Rudenko07prlb} and scattering at the nucleus \cite{Staudte07prl}.

\begin {figure}[h]
  \begin{center}
    \includegraphics[width=0.9\linewidth]{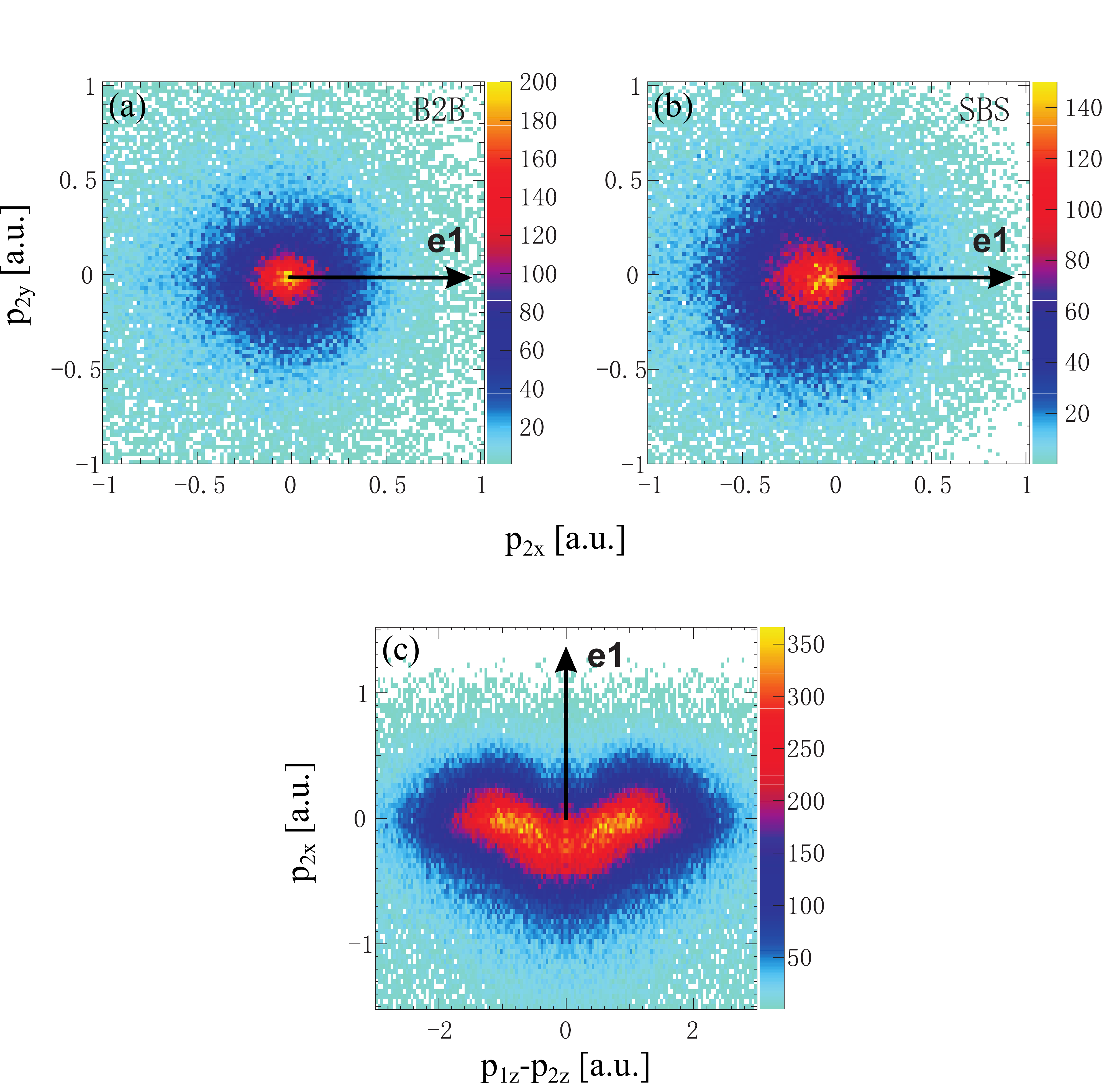}
    \caption{Momentum correlation in double ionization of He at $394$\,\unit{nm}. (a), (b) Momentum components of one electron perpendicular to the laser polarization. The horizontal axis is given by the momentum of the other electron perpendicular to the laser field, as indicated by the arrow. (a) includes events for which the momentum component of the electrons parallel to the laser polarization has opposite sign (termed "back-to-back" emission), i.e., events from the second and fourth quadrant in Fig. \ref{figpz1pz2}. (b) includes events for which the momentum component of the electrons parallel to the laser polarization has the same sign (termed "side-by-side" emission), i.e., events from the first and third quadrant in Fig. \ref{figpz1pz2}. (c) The horizontal axis shows the difference of the momentum components of the electrons parallel to the polarization, vertical axis is the same as the horizontal axis in panels (a), (b).}
  \label{figeerep}
  \end{center}
\end {figure}

Further evidence that the electron pairs with momenta in the first and third quadrant of Fig. \ref{figpz1pz2} are set free almost simultaneously during the same quarter cycle of the laser field can be gained by taking the momentum components perpendicular to the laser field into account. In this direction there is no acceleration by the light field, thus any anticorrelation between the momenta of the two electrons in this direction originates from electron-electron repulsion, i.e., occurs only if both electrons are set free simultaneously \cite{Weckenbrock03prl,Weckenbrock04prl,Weckenbrock01jpb}. Fig. \ref{figeerep}(a), \ref{figeerep}(b) show that electron pairs in the first and third quadrant of Fig. \ref{figpz1pz2} have partially opposite momentum components in the plane perpendicular to the polarization axis. In contrast, pairs from the second and fourth quadrant of Fig. \ref{figpz1pz2} show, as expected,  almost no signs of electron repulsion. 

An alternative perspective into the multidimensional momentum space highlighting the same physics point is shown in Fig. \ref{figeerep}(c) where the horizontal axis shows the momentum difference between the electrons along the polarization axis while the vertical axis shows the momentum difference perpendicular to this axis. This plot provides the most direct proof that the electrons with similar $p_z$ momentum strongly repel each other, i.e., are set free simultaneously. 

\subsection{Joint energy distributions \label{sectionjoint}}

\begin {figure}[h]
  \begin{center}
    \includegraphics[width=1.\linewidth]{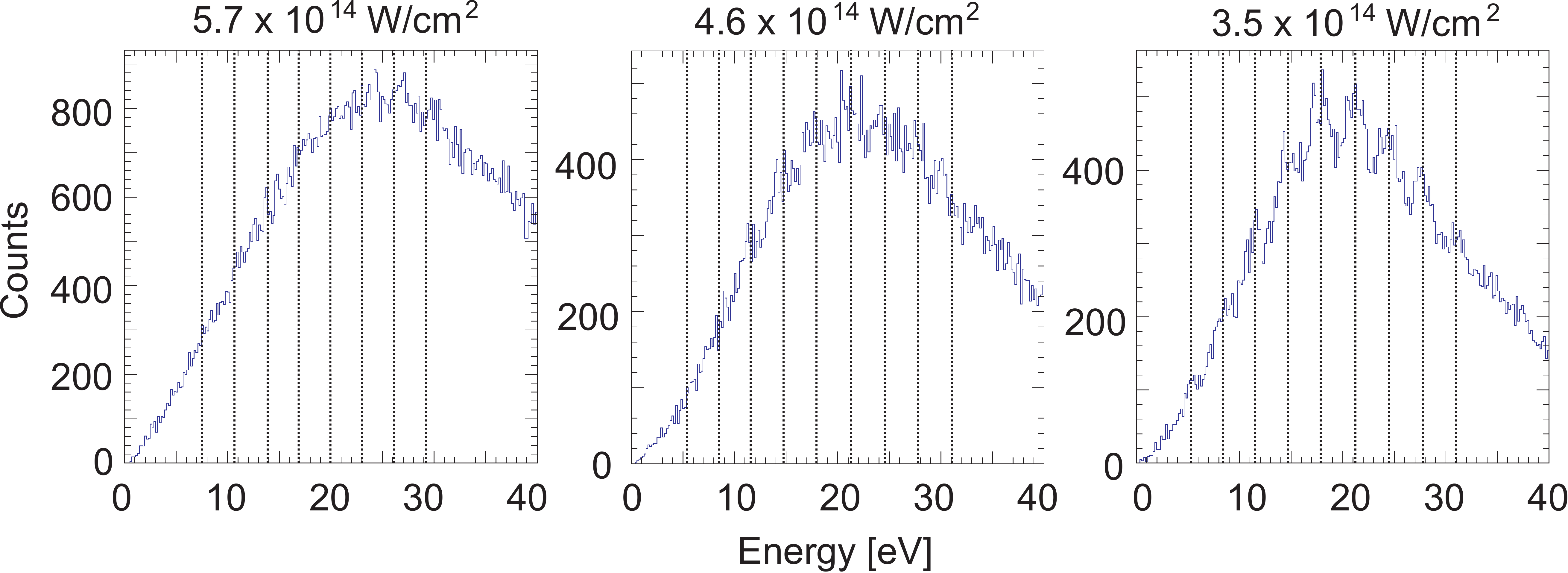}
    \caption{Distributions of the sum energy of both electrons from double ionization of He at $394$\,\unit{nm} (intensities are as given in the figure). The dashed lines show the energies predicted by equation \ref{eqneedouble}.}
  \label{figeesum}
  \end{center}
\end {figure}

The energy transfer from the laser field to the atom is quantized. This holds for single as well as for double ionization. Accordingly all direct solutions of the time-dependent Schr\"odinger equation have shown ATDI peaks as given by equation \ref{eqneedouble} \cite{Lein01pra,Liao10pra,Wang12pra,Parker06prl,Armstrong11njp,Parker01jpb,Thumm14pra,Zielinski16pra}. For Argon \cite{Henrichs13prl} we have reported experimental evidence for these structures.  We complement the equivalent representation for the two higher intensities in Fig. \ref{figeesum}. The visibility of the ATDI peaks gradually fades, which we attribute to volume averaging becoming more severe with increasing $U_p$. We note that the peaks occur at energies predicted by equation \ref{eqneedouble} which is indicated by the dashed lines. This supports the accurateness of our intensity calibration. 

\begin {figure}[h]
  \begin{center}
    \includegraphics[width=1. \linewidth]{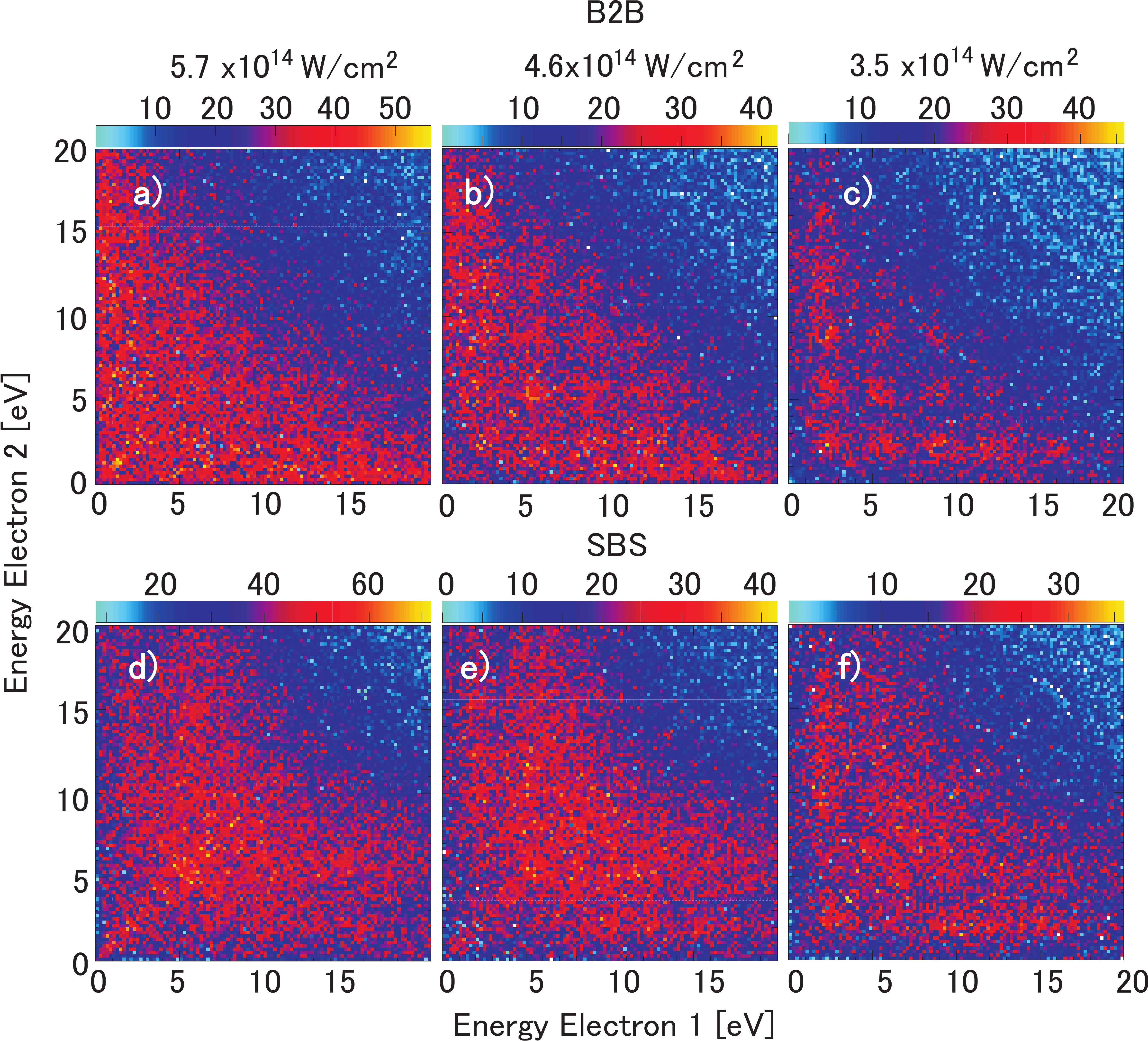}
    \caption{Joint electron energy distributions. Panels (a)-(c) show the joint energy distribution of events for which the momentum along the polarization axis has opposite sign (back-to-back (B2B), events from second and fourth quadrant of Fig. \ref{figpz1pz2}). Panels (d)-(e) contain events for which the momentum along the polarization axis has the same sign (side-by-side (SBS), events from first and third quadrant of Fig. \ref{figpz1pz2}).} 
  \label{fige1e2}
  \end{center}
\end {figure}

Equation \ref{eqneedouble} predicts only the total energy transfer to the two-electron continuum to be quantized while it leaves open, how this energy is shared between the two electrons. For the case of single photon double ionization it is known that the energy sharing is continuous \cite{Wehlitz91} with a preference towards equal energy sharing at small excess energies compared to the ionization potential and a preference towards very unequal energy sharing for very high excess energies \cite{Knapp02prl,Schoeffler13prl}. Surprisingly, for the ionization of Argon in the multi-photon regime we find that the energy sharing leads to multiple discrete peaks, i.e., that not only the sum energy but also the individual electron energy is quantized. We refer to this peaked distribution of the joint energies as "checkerboard structure". This has been confirmed for the ionization of Helium by direct solutions of the time-dependent Schr\"odinger equation \cite{Zielinski16pra}, but no explanation was given in that paper. In Fig \ref{fige1e2} we confirm the prediction at least for the lowest intensity studied here. Splitting the events into those for side-by-side (quadrant 1 and 3 in Fig \ref{figpz1pz2}) and back-to-back emission (quadrant 2 and 4 in Fig.  \ref{figpz1pz2}) shows that the checkerboard structure vanishes for electrons that are emitted to the same half sphere. This is plausible as we have seen in Fig.  \ref{figeerep} that for side-by-side emission electron repulsion becomes visible in the momentum correlation and this repulsion necessarily also leads to a continuous redistribution of energy between the electrons.

\begin {figure}[h]
  \begin{center}
    \includegraphics[width=0.9\linewidth]{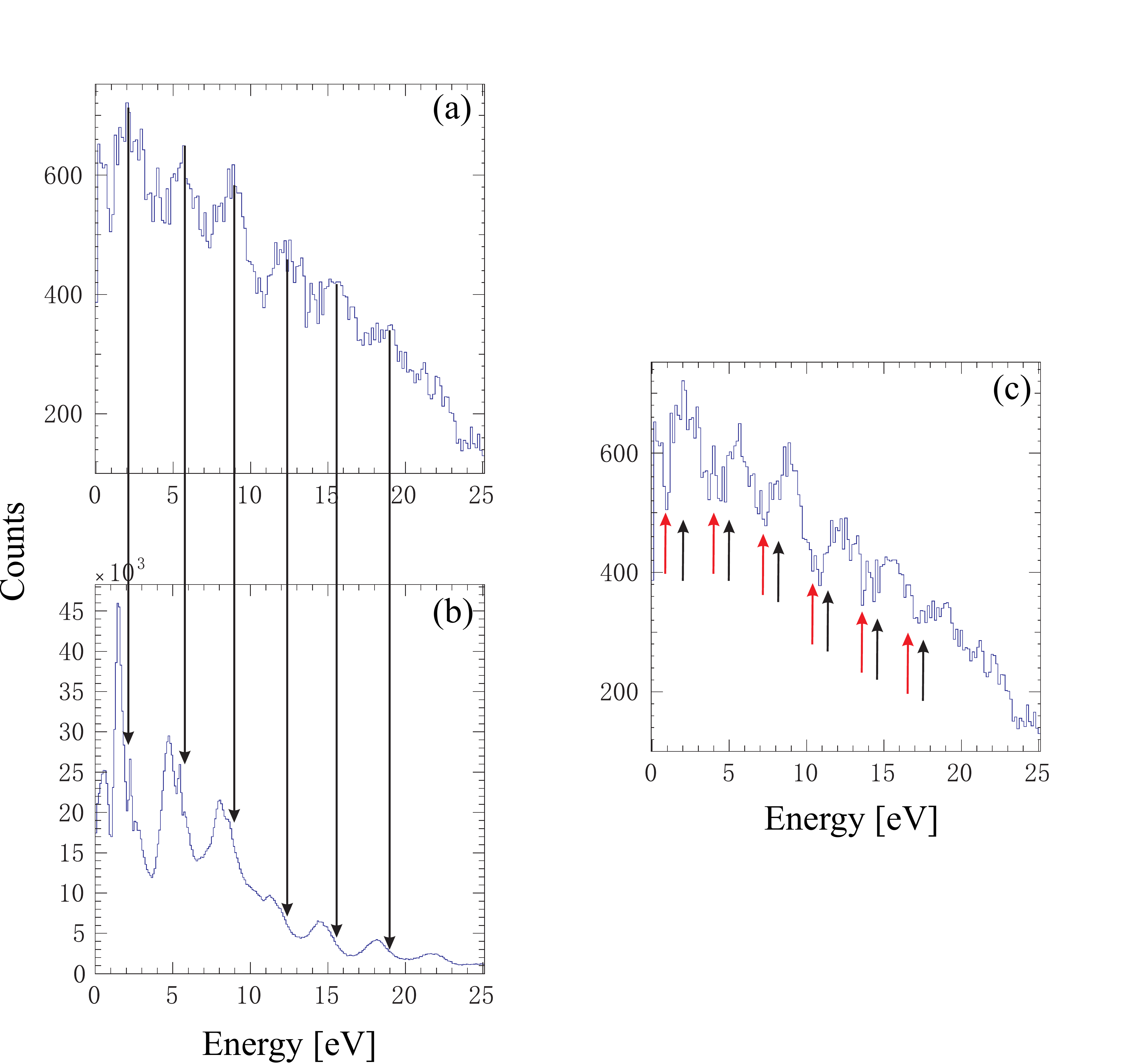}
    \caption{Comparison of peak position in electron energy distribution for single and double ionization. (a), (c) one of the electrons from double ionization for those events from Fig. \ref{fige1e2}(f). (b) single ionization. The arrows in (c) indicate the energies expected for sequential ionization calculated from Eq. \ref{eqneesingle} using $Ip_1$(black) and $Ip_2$(red).}
  \label{fige1singledouble}
  \end{center}
\end {figure}

What is the physical origin of the discretization of the single electron energies? One possible explanation could be a sequential ionization process. In this case one of the electrons (the one set free first) would show a comb of ATI peaks as in single ionization while the other electron would show peaks at energies given by equation \ref{eqneesingle} if one replaces $I_{1}$ by $I_{2}$. The test of this expectation in Fig. \ref{fige1singledouble} clearly rules out such a sequential ionization as the origin of the checkerboard structure. The location of the peaks does not coincide with the location of the peaks observed in single ionization, neither for the Freeman resonance peaks, nor for the peaks at higher energy. Furthermore the peak positions do not coincide with the prediction of equation \ref{eqneesingle} neither for $Ip_1$ (black arrows in Fig. \ref{fige1singledouble}(c)) nor for $Ip_2$ (red arrows). To unravel the origin of the checkerboard structure more theoretical work is needed.

\subsection{Joint angular distributions \label{chapterfully}}
In all figures shown so far we have integrated the data over one or several observables, such as e.g. momentum components or angles. In this section we aim for the highest level of detail avoiding integration as far as possible and analyze the data set with the lowest intensity ($3.5\times10^{14}\,W/cm^2$). To this end we select the energy of each of the electrons and the angle of one of the electrons with respect to the polarization axis and show the angular distribution for the remaining electron. This is the standard procedure studying single photon double ionization (see e.g. \cite{Briggs00jpba} for Helium and \cite{Weber04prl} for $H_2$). For multi-photon double ionization, however, this has not yet been achieved experimentally for Helium (see \cite{Henrichs13prl} for an example using a Ne target) while several theoretical studies have reported such fully differential rates \cite{Becker94pra,Thumm14pra}. For the special case of equal energy sharing and an odd number of absorbed photons the two-body continuum is shaped also by dipole selection rules. These selection rules are dealt with in \cite{Henrichs18prar} and we therefore have selected the energy regions discussed in this chapter such, that they avoid these energy regions.
\begin {figure}[h]
  \begin{center}
    \includegraphics[width=0.8\linewidth]{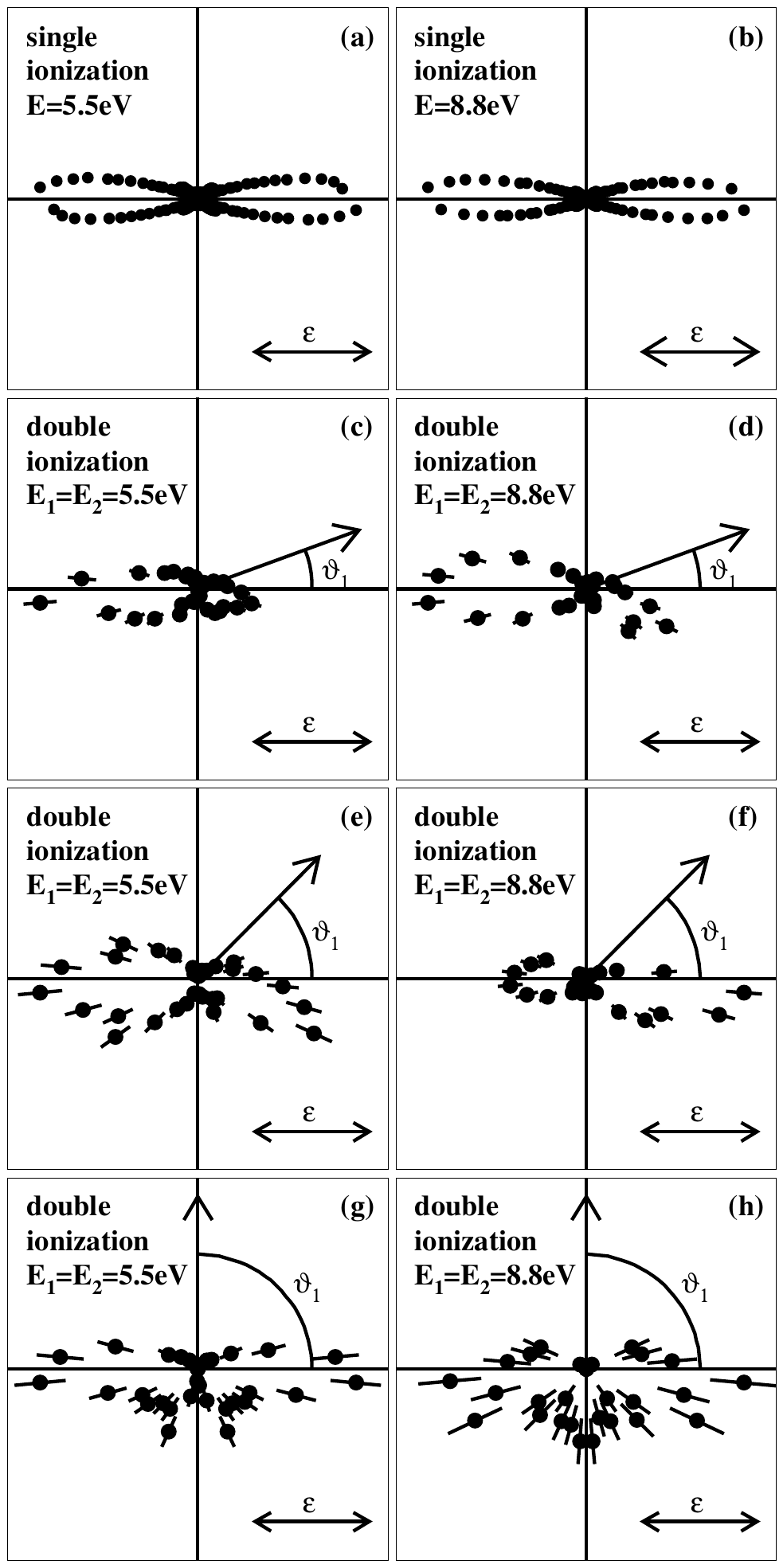}
    \caption{Joint angular distributions for double ionization, equal energy sharing. (a), (b) Electron angular distributions for single ionization of Helium at $394$\,\unit{nm} for (a) $5.5$\,\unit{eV} and $8.8$\,\unit{eV} (b) electron energy. The polarization axis is horizontal as depicted by the double arrow. (c)-(h) joint angular distributions for double ionization measured simultaneously with the data shown in (a), (b). In (c), (e), (g) only events in which both electrons have an energy of $5.5$\,\unit{eV} are selected. (d), (f), (h) show events in which both electrons have an energy of $8.8$\,\unit{eV}. The angle of the first electron is shown by the arrow. Data points show the angular distribution of electrons which are in the plane defined by the first electron and the laser's polarization. The selected energies correspond to an absorption of $32$ (c), (e), (g) and $34$ (d), (f), (h) photons.}
  \label{figfdcsequal}
  \end{center}
\end {figure}

Fig. \ref{figfdcsequal} shows joint angular distributions for equal energy sharing of both electrons. Panel (a) and (b) display the equivalent electron angular distributions for single ionization measured simultaneously. It shows the narrowly directed emission of the electron along the polarization axis. The higher energetic electrons are slightly more narrowly peaked compared to the lower energy ones. The electrons from double ionization show a similar narrowly peaked angular distribution. In all panels a clear signature of electron-electron repulsion is visible: The lobe on the right side is bend downward, avoiding the direction of the first electron. Panel (f) shows a dominance of the side-by-side emission which one might have expected from the recollision scenario where both electrons are set free simultaneously and hence driven by the laser field to the same side. For all other cases, in particular in panels (c), (d) where the first electron is selected at $20$\,deg to the polarization-axis, the relative size of the lobes is just the opposite of the one in panel (f). The side-by-side emission is strongly suppressed here. Clearly the reason is that due to electron-electron repulsion the side-by-side emission is blocked for electrons of equal energy. Thus the repulsion does bend the electrons downwards and also suppresses the emission to the same side.

\begin {figure}[h]
  \begin{center}
    \includegraphics[width=0.8\linewidth]{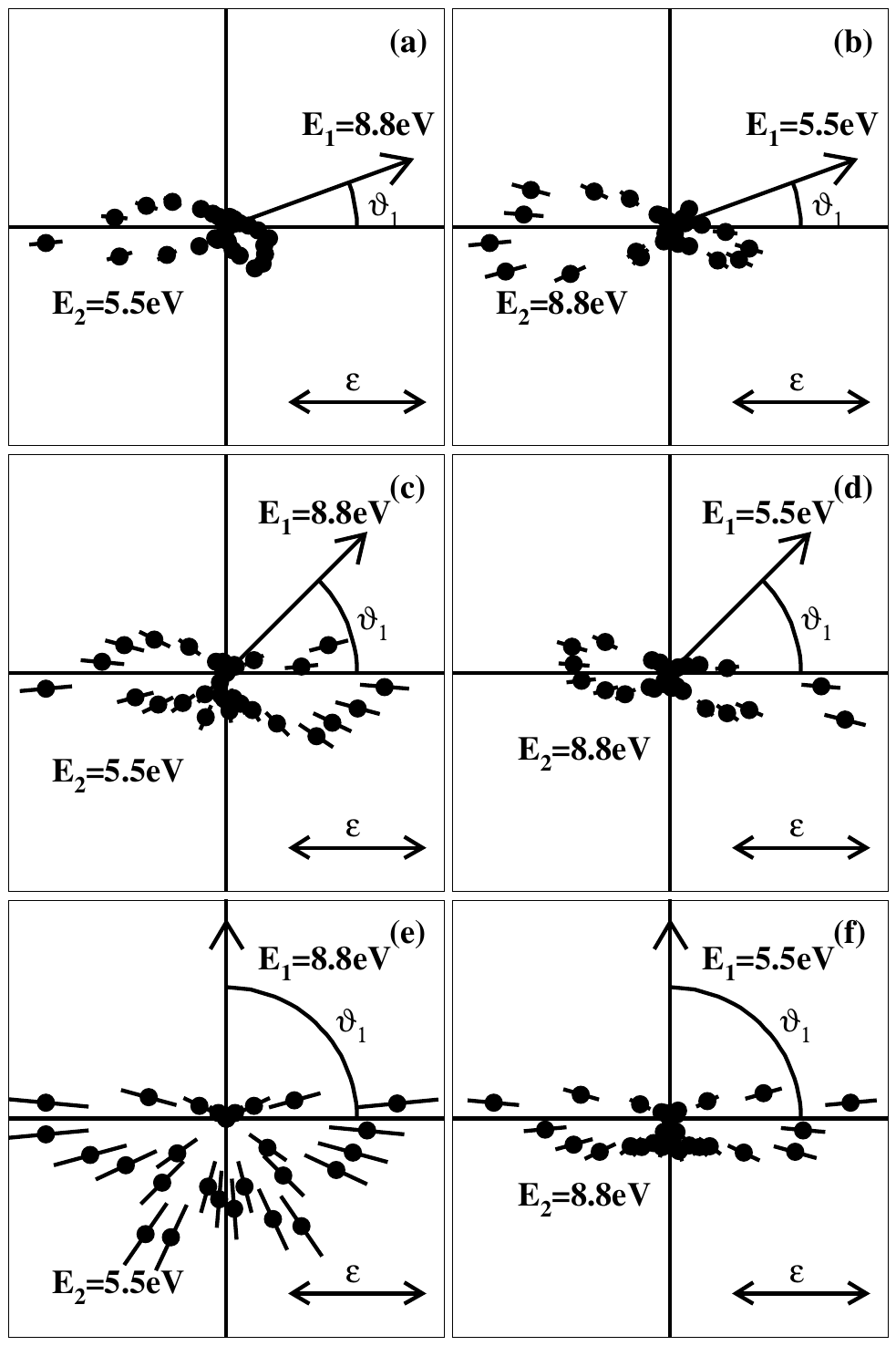}
    \caption{Joint angular distributions for double ionization, unequal energy sharing. Geometry as in Fig. \ref{figfdcsequal}, but the electrons have unequal energies as given in the panels. The selected energies correspond to an absorption of $33$ photons.}
  \label{figfdcsunequal}
  \end{center}
\end {figure}

In Fig. \ref{figfdcsunequal} we have selected unequal energies for the two electrons. The overall features of the joint angular distributions remain similar to the ones for equal energy sharing. Despite the different energies, electron repulsion still leads to a downward bend of the right lobe. Comparison of the width of the lobes shows, that the fast electron is always directed more narrowly along the polarization than the slower one. This is particular evident by comparing panels (e) and (f).

\section{Conclusion}
We have discussed a comprehensive data set on double ionization of Helium at $394$\,\unit{nm} and various intensities below the knee region of non-sequential double ionization. This study bridges between the limiting strong field tunneling case, where classical modeling of double ionization has been extremely successful and the single photon case (see also \cite{Doerner03rpa}). We find features like the emergence of a peak at zero ion momentum and peak structures in the sum electron energy as well as in the individual electron energy distributions. This demands a fully quantum mechanical modeling of the process.  The data show that electron-electron correlation plays a two-fold role in non-sequential double ionization. Firstly, it is responsible for the occurrence of double ionization, as widely recognized. Secondly, and this is a much more subtle role, it shapes the final state: momentum, energy and angular distributions as we have shown in the joint angular distributions. The latter effect is very challenging to include in theory as it requires tracking the evolution of the wave function over a very large grid. Our data can serve as a benchmark to test future calculations like the direct solution of the time-dependent Schr\"odinger equation which has become possible recently. 

\textbf{Acknowledgment} This work was supported by DFG. K. H., A. H. and K.F. thank the Studienstiftung des deutschen Volkes for financial support. H.K. was supported by the Alexander von Humboldt Stiftung. We thank A. Scrinzi, J. Zhu, V. Majety, U. Thumm and A. Becker for helpful discussions.

\bibliographystyle{unsrt}

\end{document}